\documentclass[aps,pra,twocolumn,showpacs,preprintnumbers,amsmath,amssymb,floatfix,superscriptaddress]{revtex4-1}

\usepackage{graphics}
\usepackage{graphicx}
\usepackage{amsmath}
\usepackage{setspace}
\usepackage{braket}
\usepackage{epstopdf}
\usepackage{comment}
\usepackage{hyperref}
\usepackage{bm}
\usepackage{xfrac}

\usepackage{mathrsfs}

\begin{document}


\author{Jes\'us P\'erez-R\'ios}
\address{Fritz-Haber-Institut der Max-Planck-Gesellschaft, Faradayweg 4-6, D-14195 Berlin}

\title{Cold chemistry: a few-body perspective on impurity physics of a single ion in an ultracold bath}

\date{\today}

\begin{abstract}
Impurity physics is a traditional topic in condensed matter physics that nowadays is being explored in the field of ultracold gases. Among the different classes of impurities, we focus on charged impurities in an ultracold bath. When a single ion is brought in contact with an ultracold gas it is subjected to different reactive processes that can be understood from a cold chemistry approach. In this work, we present an outlook of approaches for the dynamics of a single ion in a bath of ultracold atoms or molecules, complementing the usual many-body approaches characteristic of impurity physics within condensed matter physics. In particular, we focus on the evolution of a charged impurity in different baths, including external time-dependent trapping potentials and we explore the effect of the external laser sources present in ion-neutral hybrid traps into the lifetime of an impurity. 
\end{abstract}

\maketitle


\section{Introduction}

Cold chemistry focuses on the study of chemical reactions at temperatures 1~mK$\lesssim T\lesssim 1$~K~\cite{Krems2004,Krems2008,Schnell2009,Quemener2012,Bala2016,Carr2009,JPRBook}. At these temperatures, atoms and molecules' dynamics are presumably dominated by pure quantum mechanical behaviour, leading to intriguing phenomena on chemical reactions such as resonance effects. In addition, the average kinetic energy of atoms and molecules is comparable with the typical energy shifts caused by external fields. Thus, it is possible to control the motion and interaction of atoms and molecules efficiently. In particular, thanks to the development of ion-neutral hybrid traps, it is possible to study charged-neutral interactions in a controllable manner, paving the way to the discovery and analysis of spectacular reaction mechanisms~\cite{Hall2011,Stefan2008,Stefan2019,Saito2017,PhysRevA.101.012706,PhysRevA.91.042706,PhysRevLett.109.233202,Gianturco1,Gianturco2,RevModPhys.91.035001,Zipkes2010,Ratschbacher2012,Julienne2012,Sikorsky2018,Kleinbach2018,Ben2020,Ziv2016}.

Nowadays, ion-neutral hybrid traps serve a more ambitious purpose (apart from cold chemistry): the study of impurity physics of a charged particle in a sea of neutrals\footnote{We want to emphasize that there is an alternative to create a charged impurity in a neutral bath. This can be achieved by generating a Rydberg excitation in a Bose-Einstein condensate, which is ulterior ionized with appropriate pulse sequences~\cite{dieterle2020transport}. However, this method has one caveat: it is only useful for an ion in its parent gas.}. Impurity physics is a traditional arena in condensed matter physics, based on the concept of quasiparticle: an impurity-bath hybrid entity with particular dynamics. A prime example is Landau and Pekar's work~\cite{Landau1933,Pekar}, in which they realized that charge carriers (impurity) interact with the vibrations of the lattice (bath), dressing the charge carrier with the subsequent formation of a quasiparticle known as polaron. The polaron idea plays a crucial role in understanding the most intriguing phenomena of condensed matter physics like colossal magnetoresistance\cite{Mannella2004,Zhao2007}, charge transport in organic semiconductors\cite{Hulea2006,Podzorov2004,Watanabe2014}, and high-temperature superconductivity\cite{Mott_1993,Bussmann-Holder2007,Swartz1475}. Indeed, it has been possible to study the polaron model beyond the paradigm of material science thanks to atomic and molecular systems at ultracold temperatures\cite{Schirotzek2009,Pol1,Pol2,Pol3,Pol4,Pol5,Pol6,Pol7,Pol8,Yan190}.

 In general, the study of few-body processes, like chemical reactions (a few-body process per se), is seen as a topic disentangled from condensed matter physics. However, any many-body theory requires input from few-body processes that can constrain, affect, and ultimately control some of the system's many-body aspects. A prime example is a Rydberg impurity in an ultracold gas~\cite{Pfau2013}. In this scenario, the Rydberg atom induces collective excitations in the ultracold bath during the Rydberg atom's lifetime. However, it was observed that the Rydberg lifetime is much shorter when placed in a bath than in vacuum. This was a mystery until it was realized that the faster decay rate is a consequence of an inelastic few-body process and a reactive one~\cite{PRX2016,Whalen2017}. The inelastic process is known as $l$-changing collision~\cite{lmixing}, in which the Rydberg state ends up in a high angular momentum state after colliding with a neutral atom. The reactive process is the associative ionization~\cite{AI} (one of the chemi-ionization mechanisms~\cite{Chemi-Ionization}) in which the Rydberg electron is ionized leading to the formation of a molecular ion.

In ion-neutral hybrid traps, the longer-range nature of the charged-neutral interaction compared to neutral-neutral one may lead to new and unexpected many-body phenomena~\cite{Meso1,Massignan2005,Meso2,Meso3,astrakharchik2020ionic}. However, a more prominent long-range interaction leads inexorably to chemical reactions, which affects the state and lifetime of the impurity (if it contains internal degrees of freedom), as it has been recently shown~\cite{PRR2020,PRX2016,mohammadi2020life}. Therefore to have a comprehensive understanding of the many-body physics of a charged impurity in an ultracold bath, it is necessary to characterize the reactive and inelastic processes of the system, which, frequently, is further complicated by the presence of time-dependent trapping potentials.  
 
In this work, we present a few-body perspective into the physics of a charged impurity in an ultracold bath, emphasizing the role of cold chemical reactions required for many-body models. In particular, we highlight ion-atom-atom three-body recombination reactions relevant for a single ion in an ultracold atomic gas, which leads to the formation of weakly bound molecular ions and the formation of molecular ions from a single atomic ion in a bath of ultracold molecules, a mainly unexplored arena. Moreover, we present our particular vision of the field and what it could be in its near future, always keeping in mind that few-body and many-body physics complement each other to understand the microscopic world of atoms and molecules.

\section{A necessary preamble}
\label{Preamble}

 Before going into this article's topic, we think it is essential to emphasize that cold chemistry, as stated in the introduction, stands for any chemical reaction occurring at temperatures between 1~K and 1~mK, where quantum mechanical effects dominate different scattering observables. However, the relevance of quantum mechanical effects in a given range of temperatures depends on the underlying inter-particle interaction. For instance, let us assume that the long-range interaction between two particles is given by $-C_n/R^n$, where $n\ge 3$ and $C_n$ is the long-range coefficient. The number of partial waves, $l_{\text{Scatt}}$, contributing to any scattering observable at given collision energy, $E_k$, is given by\footnote{The highest partial wave contribution to the scattering is obtained by equating the height of the centrifugal barrier to the collision energy.}

\begin{equation}
\label{eq1}
l_{\text{Scatt}}=\left(\frac{2}{n-2}\right)^{\frac{n-2}{2n}}\sqrt{n\mu}C_n^{1/n}E_k^{\frac{n-2}{2n}},
\end{equation}

\noindent
where $\mu$ is the reduced mass of the colliding partners. Therefore, the inter-particle interaction potential sets the number of partial waves relevant for scattering observables at a given collision energy.

\begin{figure}[h]
\begin{center}
 \includegraphics[width=1\linewidth]{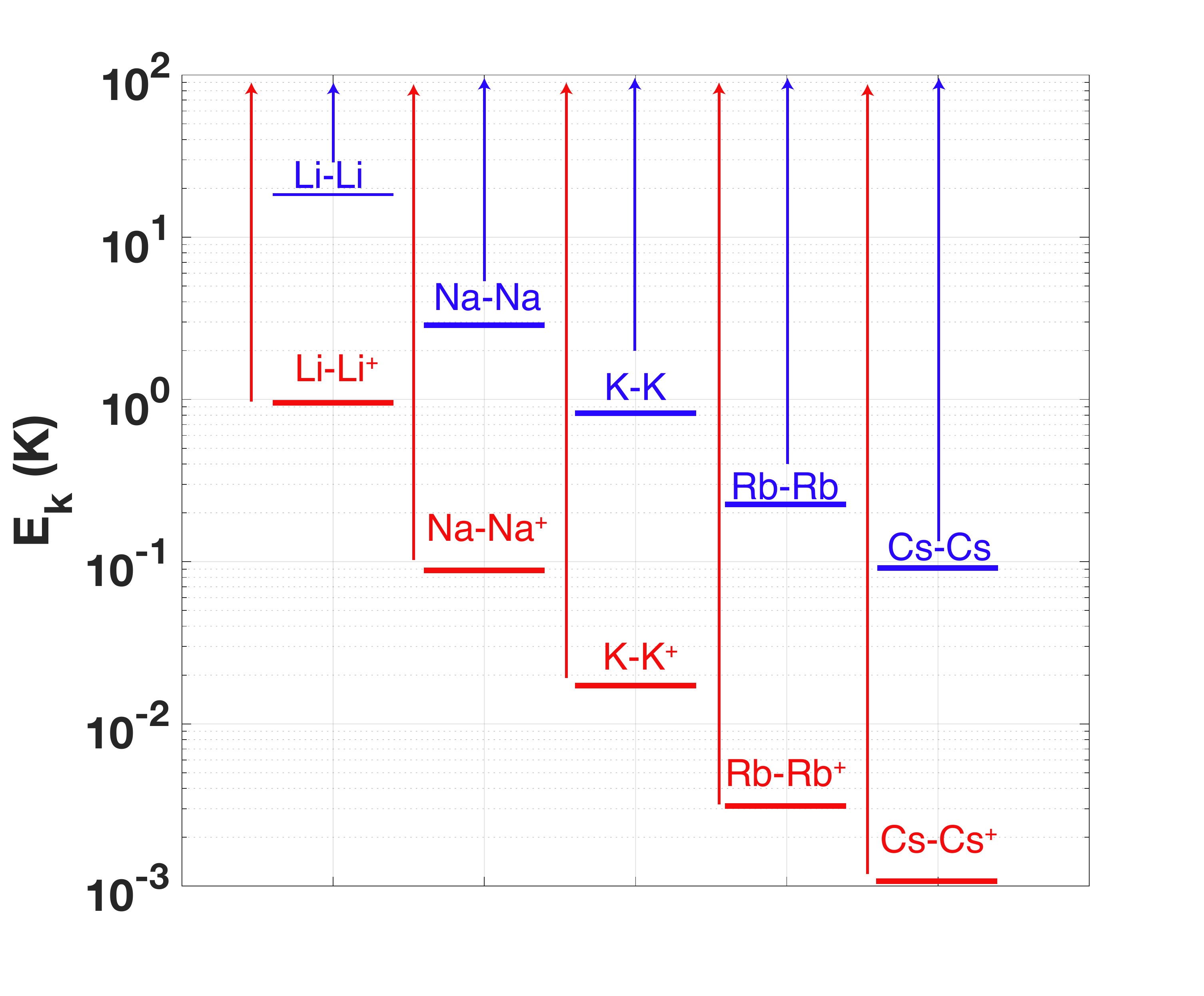}  
\caption{\label{Fig1} Lower collision energy at which 20 partial waves contribute to the scattering observable. The blue lines refer to neutral-neutral interaction, whereas the red lines are the same magnitude but for neutral-charged interactions. Figure adapted from Ref.~\cite{JPR2018}. }
\end{center}
\end{figure}

Assuming that at $l_{\text{Scatt}}\sim 20$ quantum mechanical effects are probably washed out, we have computed the lowest collision energy at which a classical approach is reliable [through Eq.~(\ref{eq1})] and the results for alkali-alkali and alkali-alkali ion collisions are presented in Fig.~\ref{Fig1}. As a result, we notice that most of the cold and ultracold chemistry processes in ion-neutral hybrid traps, involving ions and atoms, are well-suited scenarios for a classical trajectory approach~\cite{JPR2019,JPRBook}.

\section{An atomic ion in a bath of ultracold atoms}

A single ion trapped in a sea of ultracold atoms is the prototype arena for the study of exotic impurity physics, and it has been extensively studied mainly from a many-body perspective~\cite{Meso1,Massignan2005,Meso2,Meso3,astrakharchik2020ionic}. As a result, it is predicted that a single ion in a bath of ultracold atoms may form a bound state with the surrounding atoms leading to the formation of a ``mesoscopic molecular ion"~\cite{Meso1,Meso2,Meso3,astrakharchik2020ionic}. Most of the literature about this topic focuses on the system's ground state properties instead of its dynamics, with the exception of the work of Schurer et al.~\cite{Meso2}, in which the motion of the ion in a harmonic trap is included. 


An ion, $\text{A}^+$, in contact with a high density gas of ultracold atoms experiences three-body collisions with the atoms of the bath, B, leading to the formation of molecular ions
 
\begin{eqnarray}
\label{eq2}
\text{A}^+ + \text{B} + \text{B} &\rightarrow & \text{AB}^+ + \text{B},
\end{eqnarray}

\noindent
or neutral molecules

\begin{eqnarray}
\label{eq3}
\text{A}^+ + \text{B} + \text{B} &\rightarrow & \text{B}_2 + \text{A}^+,
\end{eqnarray}
 
\noindent
through ion-atom-atom three-body recombination. In three-body collisions involving charged and neutral particles, the charged-neutral interaction surpasses the neutral-neutral interaction in the energetic pooling towards three-body recombination, which translates into two important features. The first is that molecular ions are preferentially formed in comparison to neutral molecules~\cite{JPR2015,JPR2018,dieterle2020transport}, as it is sketched on panel (i) of Fig.~\ref{Fig4}. In other words, the reaction channel~(\ref{eq2}) is dominant for ion-atom-atom collisions. The preponderance of molecular ions as the final product state has been experimentally confirmed ~\cite{PRL2016}, and partially supported through full quantum mechanical calculations regarding the formation of molecular anions in some systems~\cite{Wang2017,Wang2019}. Therefore, it seems that these investigations at the cold regime may be extensible, under proper circumstances, to the ultracold realm. The second is that the ion-atom-atom three-body recombination rate shows a threshold behavior as a function of the collision energy as ~\cite{JPR2015}

\begin{equation}
\label{eq4}
k_{3}(E_k)\propto E_k^{-3/4},    
\end{equation}

\noindent
where $E_k$ denotes the collision energy. Eq.~(\ref{eq4}) has been derived assuming a classical trajectory approach in hyperspherical coordinates~\cite{JPR2014}, which is well suited to study the dynamics of atom-ion systems at cold temperatures as explained in Section~\ref{Preamble}. In particular, the motion of the nuclei is governed by the electronic potential energy surface (quantum chemistry calculations) as prescribed by Hamilton's equations. The validity of Eq.~(\ref{eq4}) has been experimentally corroborated for Ba$^+$ + Rb + Rb collisions~\cite{PRL2016} for $E_k \sim 10$~mK.

Moreover, it has been possible to derive a first-principle thermal-averaged three-body recombination rate for A$^+$ + A + A, where A is a noble gas atom, given by

\begin{eqnarray}
k_3(T) &=&\frac{1}{2(k_BT)^3}\int_{0}^{\infty}k_{3}(E_k)E_k^2e^{-\frac{E_k}{k_BT}}dE_k \nonumber \\
&=&\frac{8\pi^2}{15}\frac{\Gamma(9/4)3^{1/4}}{\sqrt{2}(k_BT)^{3/4}}\frac{(2\alpha)^{5/4}}{\sqrt{m}},    
\label{eq5}
\end{eqnarray}

\noindent
where $\alpha$ is the polarizability of the noble gas atom, $m$ is the mass of the noble gas atom, $k_B$ is the Boltzmann constant, $T$ stands for the temperature of the gas and $\Gamma(x)$ is the Euler gamma function of argument $x$.  Eq.~(\ref{eq5}) correctly describes the dependence of the three-body recombination rate as a function of the noble gas atom properties.

For three-body collisions, the collisional time can be estimated as 

\begin{equation}
\label{eq6}    
\tau=\frac{1}{\rho^2 k_3},    
\end{equation}

\noindent
where $\rho$ is the density of the gas (in the case of ion-atom-atom three-body recombination is the density of atoms). For Ba$^+$ + Rb + Rb collisions at cold temperature the three-body recombination rate is $\sim 10^{-24}$ cm$^6$/s~\cite{PRL2016} and assuming a typical atomic density of an atomic gas close to quantum degeneracy ($\rho\sim 10^{14}$~cm$^3$), the collisional time is $\sim 100\mu$s. In other words, after 100$\mu$s, in average, the ion impurity will evolve into a molecular ion, thus given an upper limit to the lifetime of the impurity and hence constraining any many-body phenomena occurring in the system.

\subsection{A weakly bound molecular ion in an ultracold bath of atoms}
\label{quenching}

The resulting molecular ion after ion-atom-atom three-body recombination appears in a weakly bound vibrational state whose binding energy correlates with the collision energy~\cite{PRL2016}. The molecular ion may further collide with atoms of the bath, thus it can be seen as a new class of impurity. The dynamics of this {\it new} impurity can be studied from a quasi-classical trajectory (QCT) perspective in which Newton's law of classical mechanics describes the motion of the nuclei in the electronic potential energy surface. In contrast, the initial conditions for the trajectories are chosen according to the colliding partners' quantum state through the Wentzel, Kramers, and Brillouin (WKB) or semi-classical approximation. The same methodology is applied to map the final values of position and momenta of the trajectories within the phase-space into quantum states of the system at hand~\cite{Karplus-1965,Truhlarbook,JPR2019,JPRBook}.

A weakly bound molecular ion colliding with an atom leads, apart from the expected elastic collisions, to three possible outcomes. First, vibrational quenching mediated by an efficient translational-vibrational energy transfer as AB$(v)^+$ + B $\rightarrow$ AB$(v')^+$ + B, with $v' \ne v$ where $v$ stands for the vibrational state of the molecular ion as sketched in panel (ii) of Fig.~\ref{Fig4}. Second, molecular formation through an exchange reaction as AB$^+$ + B $\rightarrow$ B$_2$ + A$^+$, as it is depicted in panel (iii) of Fig.~\ref{Fig4} assuming that the charge is located at the atom A and that the dynamics occur in a single potential energy surface. In other words, charge transfer reactions are not considered. Third, dissociation of the molecular ion leading to two free atoms and one ion as AB$(v)^+$ + B $\rightarrow$ A$^+$ + B + B, which is schematically shown in panel (iv) of Fig.~\ref{Fig4}. From all these possible reaction channels, the most relevant at cold temperatures is the vibrational quenching followed by the dissociation, which is only active as long as the collision energy is larger than the molecular ion's binding energy. However, the formation of neutral molecules is not relevant at cold temperatures.

\begin{figure}[h]
\begin{center}
 \includegraphics[width=0.75\linewidth]{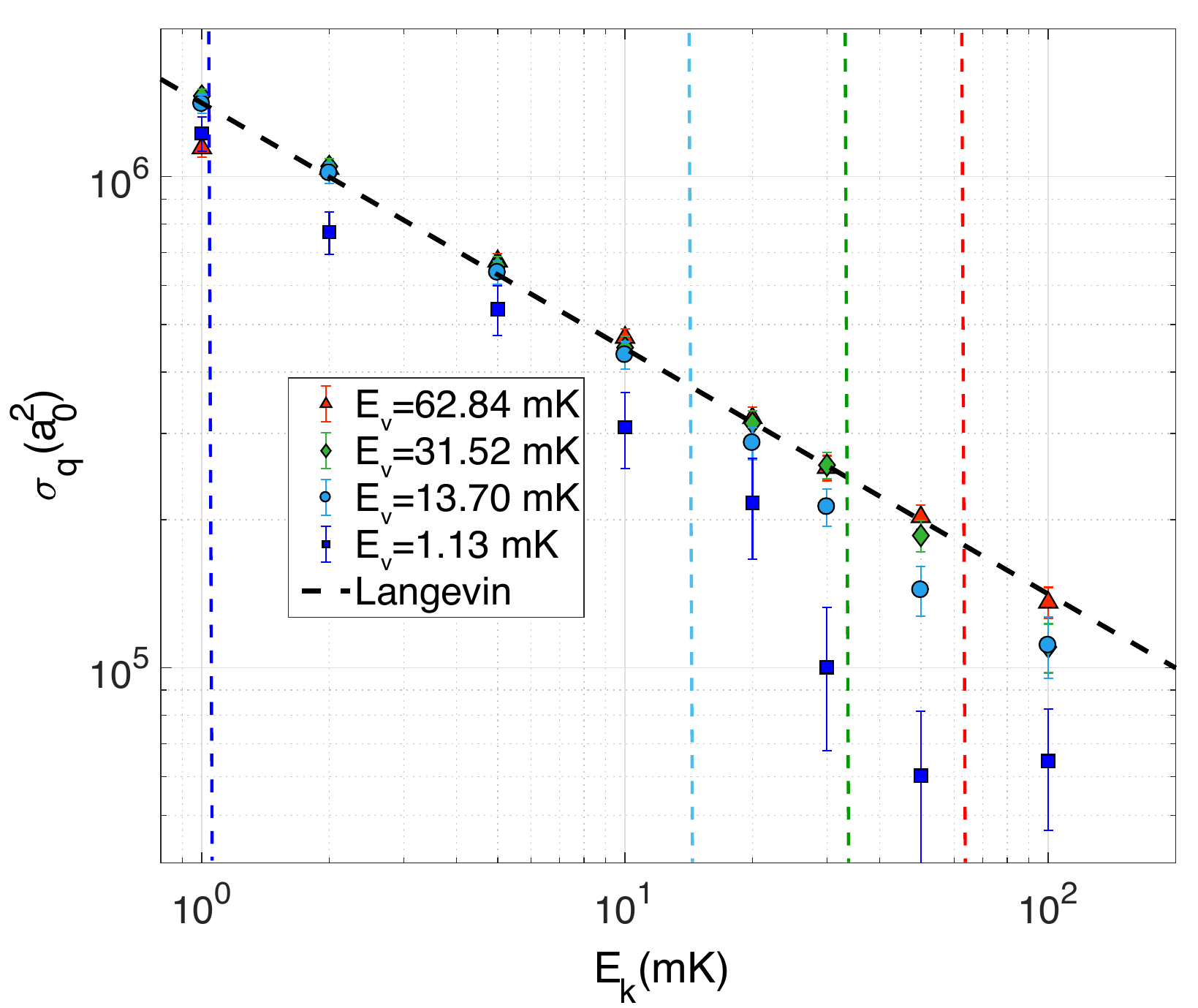}  
\caption{\label{Fig3}  Quenching cross section for the collision BaRb$^+$ ($v$) + Rb $\rightarrow$ BaRb$^+(v'\ne v)$ + Rb as a function of the collision energy ($E_k$). The different binding energies ($E_v$) of the vibrational states $v$ are denoted by different symbols as indicated on the legend. The dashed line represent the Langevin cross section. The vertical dashed lines stand for the binding energy of the initial vibrational states of the molecular ion. Figure adapted from Ref.~\cite{JPR2018}.}
\end{center}
\end{figure}

As an example, the vibrational quenching cross section computed via a QCT formalism for RbBa$^+ (v )$ + Rb as a function of the collision energy for different vibrational states is shown in Fig.~\ref{Fig3}. In this figure, it is noticed that the vibrational quenching cross section is mostly independent of the vibrational state of the molecular ion, and its tendency agrees with the predictions based on the  Langevin capture model, $\sigma_L(E_k)=\pi\sqrt{2\alpha/E_k}$~\cite{Langevin}, which is represented as the black dashed line. The regions where the QCT results deviate from the Langevin prediction correspond to collision energies larger than the molecular ion's binding energy, denoted as the vertical dashed lines in Fig.~ \ref{Fig3}. Indeed, the agreement between the vibrational quenching cross section and the Langevin prediction translates into a very efficient vibrational-translational energy, which can be rationalized in terms of the adiabaticity parameter~\cite{JPRBook,JPR2019,Levine}.

\subsection{The role of external laser sources}

Every single ion-neutral hybrid trap experiment requires different laser sources to laser-cool ions and to trap, hold and manipulate ultracold atoms. These laser sources, are generally assumed to have a little impact on the dynamics of a single ion in a bath of ultracold atoms. However, it has been shown that the trapping lasers of a magneto-optical trap holding ultracold atoms may lead to an enhancement of the rate of charge-transfer reactions~\cite{Hall2011,Hall2013,Hall2013bis,Ratschbacher2012}. Therefore, laser sources may play a relevant role on the study of a charged impurity in a bath of ultracold atoms.

\begin{figure}[h!]
\begin{center}
 \includegraphics[width=1\linewidth]{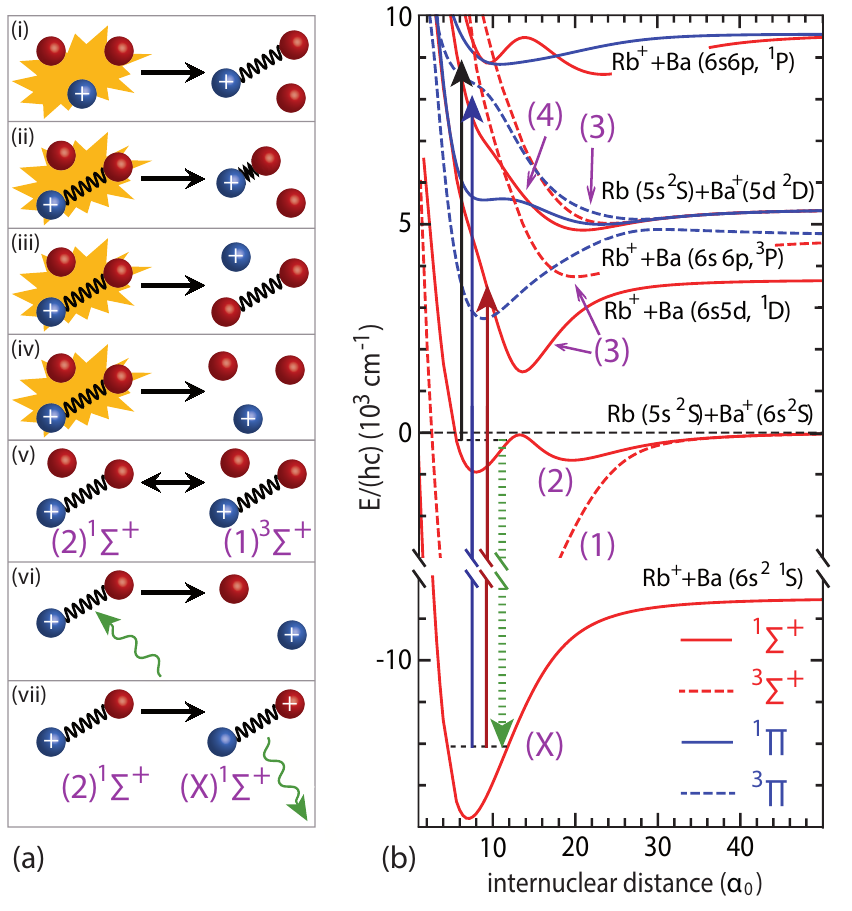}  
\caption{\label{Fig4} Collision channels (radiative and non-radiative) of a weakly bound molecular ion in a high density ultracold gas. Panel (i) represents ion-atom-atom three-body recombination; panel (ii) stands for a vibrational quenching of a molecular ion, panel (iii) is a substitution chemical reaction in which a molecule appears as the product state; panel (iv) represent a dissociation process, panel (v) is a spin-flip collision, panel (vi) is associated with a photo-dissociation process; and panel (vii) stands for a radiative decay process. The right panel shows the potential energy curves relevant for the different radiative-assisted process in BaRb$^+$. Solid black, blue, and red arrows show possible photodissociation transitions for 1064 nm, 493 nm, and 650 nm light, respectively. The dashed green arrow indicates radiative relaxation to the electronic ground state. Figure taken from Ref.~\cite{mohammadi2020life}.}
\end{center}
\end{figure}

A prime example of this phenomenology is a single ion immersed in a high density ultracold atomic gas~\cite{mohammadi2020life}. In this scenario, as explained above, a weakly bound molecular ion emerges due to an ion-atom-atom three-body recombination reaction. The resulting weakly bound molecular ion may be dissociated by the external light sources, as it is shown in panels (vi) and (b) of Fig.~\ref{Fig4}. In this figure, it is shown the different processes and relevant potential energy curves that play a role in the dynamics of a weakly bound molecular ion, BaRb$^+$, in a sea of ultracold Rb atoms (the black dashed line below the origin of energy represents the vibrational state of the molecular ion). The trapping light (in this case of 1064 nm is represented as the black arrow) couples weakly vibrational sates of  BaRb$^+$ to dissociative electronic states inducing its photo-dissociation. However, photo-dissociation needs to compete with vibrational quenching, dissociation and substitution chemical reaction, introduced in Sec.~\ref{quenching} and depicted in panels (ii-iv), as well as with spin-flip transitions, which may transfer the molecular ion into the metastable triplet electronic state, $(1)^3\Sigma^+$. Moreover, the weakly bound molecular ion may decay into a highly vibrational state of the (X)$^1\Sigma^+$ ground electronic state via spontaneous emission. Finally, the molecular ion can relax into deeply bound vibrational states thanks to vibrational quenching collisions or radiative decay processes, and finally, it can be photo-dissociated by the lasers employed to laser-cool Ba$^+$. The role of each of the radiative and non-radiative processes has been studied using QCT results for the vibrational quenching, dissociation and substitution reactions, a semi-classical approach for the spin-flip transitions and a full-quantum treatment of the photo-dissociation cross section~\cite{mohammadi2020life}. This complex cold reaction network is experimentally corroborated by looking at the relevance of different ionic species as a function of time~\cite{mohammadi2020life}. Finally, we would like to point out that the role of external laser sources on few-body processes has also been observed in molecular ion-atom systems~\cite{Stefan2019,Sourav2016}.

\section{An atomic ion in a bath of molecules}

The study of an atomic ion's dynamics in an ultracold bath of molecules has received little attention. Indeed, only recently, Hirzler et al. have addressed the dynamics of a single ion colliding with ultracold molecules~\cite{PRR2020}. In that work, the dynamics of a single Yb$^+$ in a bath of ultracold molecules is studied within a QCT approach, including the time-dependent trapping potential of an ion in a Paul trap. The bath consists of a gas of ultracold Li$_2$ Feshbach molecules whose binding energy can be tuned through an external magnetic field. A single ion in a bath of molecules shows three different reaction pathways, which are depicted in Fig.~\ref{Fig5}:

\begin{figure}[h]
\begin{center}
 \includegraphics[width=0.75\linewidth]{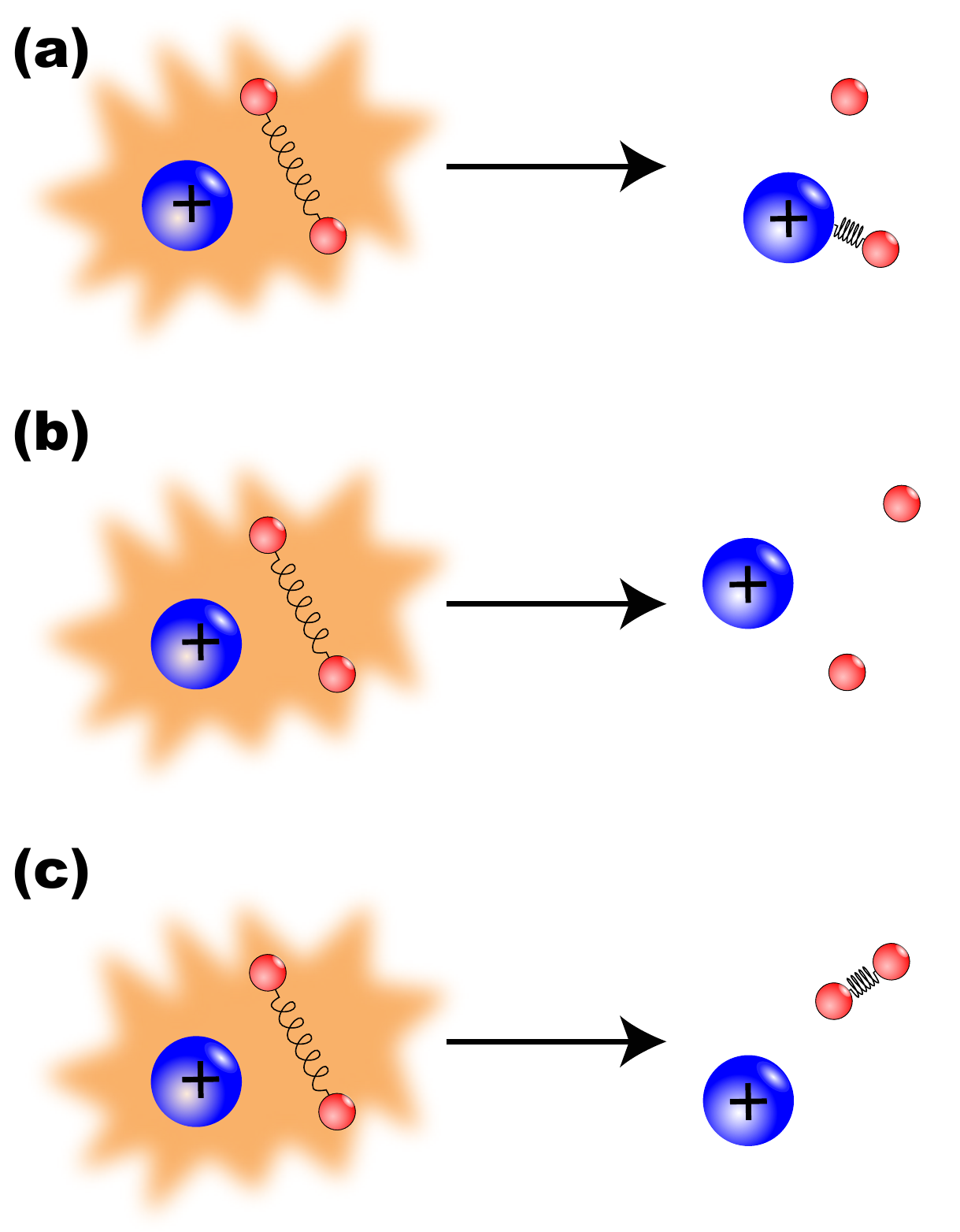}  
\caption{\label{Fig5} A sketch of the dynamics of an atomic ion colliding with a molecule. Panel (a) stands for the molecular ion formation process, panel (b) for the dissociation of the molecule and panel (c) represents the vibrational quenching of the molecule.  }
\end{center}
\end{figure}

\begin{itemize}
    \item Molecular ion formation: Yb$^+$ + Li$_2$ $\rightarrow$ YbLi$^+$ + Li.
    \item Vibrational quenching of the molecule:  Yb$^+$ + Li$_2(v)$ $\rightarrow$ Yb$^+$ + Li$_2(v')$ with $v\ne v'$. 
    \item Molecular dissociation: Yb$^+$ + Li$_2$ $\rightarrow$ Yb$^+$ + Li + Li.
\end{itemize}

\noindent
Among them, molecular ion formation and molecular dissociation are the most relevant processes depending on the collision energy and binding energy of the molecule, whereas vibrational quenching is irrelevant for typical conditions in ion-neutral hybrid traps.

\begin{figure}[h!]
\begin{center}
 \includegraphics[width=1\linewidth]{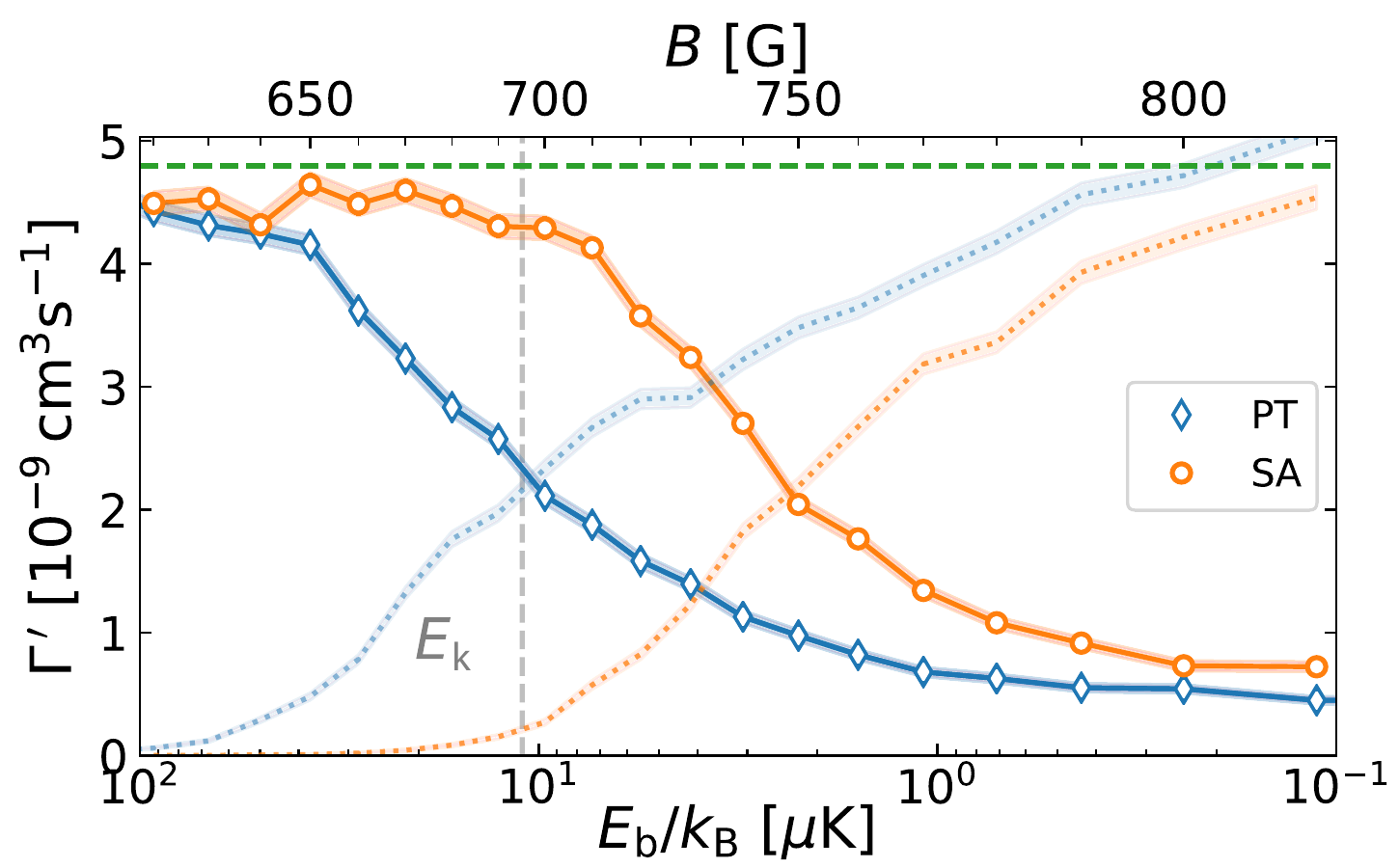}  
\caption{\label{Fig6}  Reaction rates for molecular ion formation (solid lines) and dissociation (dotted lines) as a function of the binding energy of the molecule for a collision energy $E_k \approx  11 \mu$K (gray dashed line), corresponding to $T_{\text{Li}_2}=2\mu$K and $T_{\text{Yb}^+} =100\mu$K. The results including explicitly the time-dependent trapping potential are labeled as PT, whereas the results within the time-independent secular approximation are labeled as SA. The green dashed line is the Langevin collision rate for Li-Yb$^+$. Figure adapted from Ref.~\cite{PRR2020}}
\end{center}
\end{figure}

The reaction rate for molecular ion formation and dissociation for Yb$^+$ + Li$_2$ for fixed collision energy, $E_k$, as a function of the binding energy of the Li$_2$ molecule is shown in Fig.~\ref{Fig5}. Each reaction rate has two results: one corresponds to the results assuming the time-independent secular approximation (SA). The other includes the time-dependent trapping potential explicitly that the ion feels in the Paul trap (PT). In this figure, it is noticed that for tightly bound molecules $E_b\gg E_k$, the ion reacts with the molecule to form a molecular ion with a rate similar to the Langevin prediction. Nevertheless, for weakly bound molecules, molecular dissociation dominates the dynamics. It is also worth pointing out that the inclusion of the time-dependent trapping potential leads systematically to a lower molecular ion formation rate compared with the SA results, translating into a larger dissociation rate. 

\begin{figure}[h]
\begin{center}
 \includegraphics[width=1\linewidth]{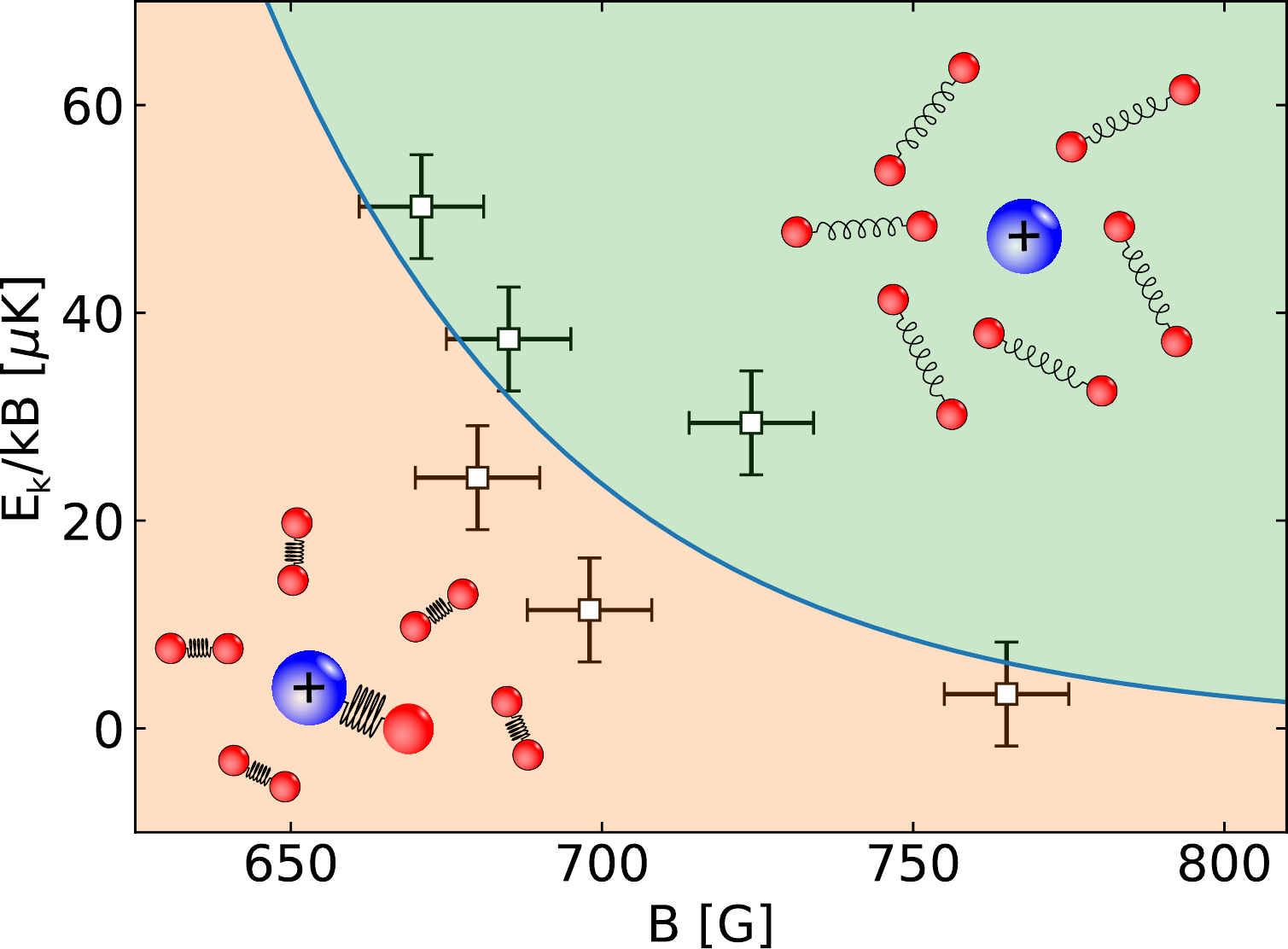}  
\caption{\label{Fig7} The phase-diagram of a single ion in an ultracold bath of weakly bound molecules. The magnetic field establishes the binding energy of the molecules in the bath through a Feshbach resonances whereas the collision energy is determined by the translational temperature of the ion in the trap. For higher ion temperatures the ion survives in the bath whereas at lower ion temperature the ion reacts with a molecule leading to the formation of a stable molecular ion. The error bars are characteristic of using a Monte Carlo approach for sampling the phase-space. Figure courtesy of Henrik Hirzler.}
\end{center}
\end{figure}

Exploring the dynamics of Yb$^+$ + Li$_2$ collisions for different collision energies and magnetic fields, it is possible to elaborate a ``phase-diagram" of a single ion in a bath of weakly bound molecules. As a result, it is found, as anticipated in Fig.~\ref{Fig6}, that the nature of the impurity depends on the binding energy of the molecules of the bath (controlled via the magnetic field) and on the collision energy (mainly controlled by the energy of the ion). In particular, an atomic ion impurity evolves into a molecular ion at low collision energies and considerable molecular binding energies (smaller magnetic fields). Therefore, a single ion in a bath of weakly bound molecules can be tuned into a molecular ion in a bath of weakly bound molecules, which opens up a new avenue for impurity physics, bringing (maybe) new polaronic effects.

\section{Summary and concluding remarks}

In summary, we have highlighted different scenarios in which a single ion in an ultracold bath evolves differently depending on the bath's properties, the presence of external laser sources, and time-dependent potentials. In particular, we have presented the most relevant few-body processes within cold chemistry that play a role in the dynamics and evolution of a charged impurity in an ultracold bath of atoms or molecules. As a result, a single ion in a high-density bath experiences an ion-atom-atom three-body recombination reaction leading to the formation of a weakly bound molecular ion. This molecular ion experiences two-body collisions with the bath's atoms, leading to its vibrational relaxation and ulterior photo-dissociation by the laser holding the ultracold atoms. On the other hand, a single ion in a bath of ultracold molecules will transition into a molecular ion if the bath molecules are tightly bound, whereas the ion will remain unperturbed in the case of weakly bound molecules. Indeed, it has been possible to draw a ``phase-diagram'' of an ion in a bath of Feshbach molecules, which shows the nature of the impurity as a function of the ion energy and applied magnetic field.

This article's primary goal is to emphasize that few-body processes are necessary to understand many-body phenomena in ion-neutral hybrid traps. Therefore, a full characterization of the dynamics and main properties of a charged impurity in a sea of ultracold atoms requires, in our humble opinion, a hybrid approach in which a many-body methodology incorporates few-body processes. In other words, it is necessary to reduce the human-made gap between few-body and many-body physics to understand impurity physics in ion-neutral hybrid traps properly.

\bibliography{interacttfosample}

\end{document}